
%


\magnification = 1000
\overfullrule=0pt       
\nopagenumbers          
\baselineskip=13pt      
\vsize=22.1truecm
\hsize=14.5truecm
\parindent=1.2truecm
\parskip 0pt
\def\hangsrw{\hangindent 10pt}
\def\hb{\hfill\break}
\def\ref#1{\par\hangsrw\noindent{#1\enspace}\ignorespaces}
\def\cret{\par\noindent}
\def\longindentation{\quad\qquad\qquad\qquad\qquad\qquad\qquad\qquad}
\def\ie{{\it i.e.,\ }}
\def\eg{{\it e.g.,\ }}
\tolerance = 3000


\def\beginparmode{\endmode
  \begingroup \def\endmode{\par\endgroup}}
\let\endmode=\par
\def\references{\def\endmode{\r@ferr\par\endgroup}}
\def\endreferences{}

\def\references                 
  {
   \frenchspacing \parindent=0pt
   \parskip=0pt
   \everypar{\hangindent=0.65truecm}
  }

\def\ref#1{Ref.~#1}                     
\def\Ref#1{Ref.~#1}                     
\def\[#1]{[\cite{#1}]}
\def\cite#1{{#1}}
\def\(#1){(\call{#1})}
\def\call#1{{#1}}
\def\taghead#1{}
\def\frac#1#2{{#1 \over #2}}
\def\half{{\frac 12}}
\def\third{{\frac 13}}
\def\fourth{{\frac 14}}
\def\12{{1\over2}}
\def\eg{{\it e.g.,\ }}
\def\Eg{{\it E.g.,\ }}
\def\ie{{\it i.e.,\ }}
\def\Ie{{\it I.e.,\ }}
\def\etal{{\it et al. \ }}
\def\etc{{\it etc.\ }}
\def\via{{\it via\ }}
\def\cf{{\sl cf.\ }}

\def\refto#1{$^{#1}$}           

\gdef\refis#1{#1.\hskip0.4truecm}                     

\gdef\journal#1, #2, #3, 1#4#5#6{         
    #1~{\bf #2}, #3 (1#4#5#6)}            

\def\refstylenp{                
  \gdef\refto##1{ [##1]}                        
  \gdef\refis##1{\item{##1)\ }}                 
  \gdef\journal##1, ##2, ##3, ##4 {             
     ##1~{\bf ##2~}(##3) ##4 }}

\def\refstyleprnp{              
  \gdef\refto##1{ [##1]}                        
  \gdef\refis##1{\item{##1)\ }}                 
  \gdef\journal##1, ##2, ##3, 1##4##5##6{       
    ##1~{\bf ##2~}(1##4##5##6) ##3}}

\def\pr{\journal Phys. Rev., }

\def\pra{\journal Phys. Rev. A, }

\def\prb{\journal Phys. Rev. B, }

\def\prc{\journal Phys. Rev. C, }

\def\prd{\journal Phys. Rev. D, }

\def\prl{\journal Phys. Rev. Lett., }

\catcode`@=11
\newcount\r@fcount \r@fcount=0
\newcount\r@fcurr
\immediate\newwrite\reffile
\newif\ifr@ffile\r@ffilefalse
\def\w@rnwrite#1{\ifr@ffile\immediate\write\reffile{#1}\fi\message{#1}}

\def\writer@f#1>>{}
\def\referencefile{
  \r@ffiletrue\immediate\openout\reffile=\jobname.ref%
  \def\writer@f##1>>{\ifr@ffile\immediate\write\reffile%
    {\noexpand\refis{##1} = \csname r@fnum##1\endcsname = %
     \expandafter\expandafter\expandafter\strip@t\expandafter%
     \meaning\csname r@ftext\csname r@fnum##1\endcsname\endcsname}\fi}%
  \def\strip@t##1>>{}}
\let\referencelist=\referencefile

\def\citeall#1{\xdef#1##1{#1{\noexpand\cite{##1}}}}
\def\cite#1{\each@rg\citer@nge{#1}}	

\def\each@rg#1#2{{\let\thecsname=#1\expandafter\first@rg#2,\end,}}
\def\first@rg#1,{\thecsname{#1}\apply@rg}	
\def\apply@rg#1,{\ifx\end#1\let\next=\relax
\else,\thecsname{#1}\let\next=\apply@rg\fi\next}

\def\citer@nge#1{\citedor@nge#1-\end-}	
\def\citer@ngeat#1\end-{#1}
\def\citedor@nge#1-#2-{\ifx\end#2\r@featspace#1 
  \else\citel@@p{#1}{#2}\citer@ngeat\fi}	
\def\citel@@p#1#2{\ifnum#1>#2{\errmessage{Reference range #1-#2\space is bad.}
    \errhelp{If you cite a series of references by the notation M-N, then M and
    N must be integers, and N must be greater than or equal to M.}}\else%
 {\count0=#1\count1=#2\advance\count1 by1\relax\expandafter\r@fcite\the\count0,
  \loop\advance\count0 by1\relax
    \ifnum\count0<\count1,\expandafter\r@fcite\the\count0,%
  \repeat}\fi}

\def\r@featspace#1#2 {\r@fcite#1#2,}	
\def\r@fcite#1,{\ifuncit@d{#1}		
    \expandafter\gdef\csname r@ftext\number\r@fcount\endcsname%
    {\message{Reference #1 to be supplied.}\writer@f#1>>#1 to be supplied.\par
     }\fi%
  \csname r@fnum#1\endcsname}

\def\ifuncit@d#1{\expandafter\ifx\csname r@fnum#1\endcsname\relax%
\global\advance\r@fcount by1%
\expandafter\xdef\csname r@fnum#1\endcsname{\number\r@fcount}}

\let\r@fis=\refis			
\def\refis#1#2#3\par{\ifuncit@d{#1}
    \w@rnwrite{Reference #1=\number\r@fcount\space is not cited up to now.}\fi%
  \expandafter\gdef\csname r@ftext\csname r@fnum#1\endcsname\endcsname%
  {\writer@f#1>>#2#3\par}}

\def\r@ferr{\endreferences\errmessage{I was expecting to see
\noexpand\endreferences before now;  I have inserted it here.}}
\let\r@ferences=\references
\def\references{\r@ferences\def\endmode{\r@ferr\par\endgroup}}

\let\endr@ferences=\endreferences
\def\endreferences{\r@fcurr=0
  {\loop\ifnum\r@fcurr<\r@fcount
    \advance\r@fcurr by 1\relax\expandafter\r@fis\expandafter{\number\r@fcurr}%
    \csname r@ftext\number\r@fcurr\endcsname%
  \repeat}\gdef\r@ferr{}\endr@ferences}


\let\r@fend=\endpaper\gdef\endpaper{\ifr@ffile
\immediate\write16{Cross References written on []\jobname.REF.}\fi\r@fend}

\catcode`@=12

\citeall\refto		
\citeall\ref		%
\citeall\Ref		%

\catcode`@=11
\newcount\tagnumber\tagnumber=0

\immediate\newwrite\eqnfile
\newif\if@qnfile\@qnfilefalse
\def\write@qn#1{}
\def\writenew@qn#1{}
\def\w@rnwrite#1{\write@qn{#1}\message{#1}}
\def\@rrwrite#1{\write@qn{#1}\errmessage{#1}}

\def\taghead#1{\gdef\t@ghead{#1}\global\tagnumber=0}
\def\t@ghead{}

\expandafter\def\csname @qnnum-3\endcsname
  {{\t@ghead\advance\tagnumber by -3\relax\number\tagnumber}}
\expandafter\def\csname @qnnum-2\endcsname
  {{\t@ghead\advance\tagnumber by -2\relax\number\tagnumber}}
\expandafter\def\csname @qnnum-1\endcsname
  {{\t@ghead\advance\tagnumber by -1\relax\number\tagnumber}}
\expandafter\def\csname @qnnum0\endcsname
  {\t@ghead\number\tagnumber}
\expandafter\def\csname @qnnum+1\endcsname
  {{\t@ghead\advance\tagnumber by 1\relax\number\tagnumber}}
\expandafter\def\csname @qnnum+2\endcsname
  {{\t@ghead\advance\tagnumber by 2\relax\number\tagnumber}}
\expandafter\def\csname @qnnum+3\endcsname
  {{\t@ghead\advance\tagnumber by 3\relax\number\tagnumber}}

\def\equationfile{%
  \@qnfiletrue\immediate\openout\eqnfile=\jobname.eqn%
  \def\write@qn##1{\if@qnfile\immediate\write\eqnfile{##1}\fi}
  \def\writenew@qn##1{\if@qnfile\immediate\write\eqnfile
    {\noexpand\tag{##1} = (\t@ghead\number\tagnumber)}\fi}
}

\def\callall#1{\xdef#1##1{#1{\noexpand\call{##1}}}}
\def\call#1{\each@rg\callr@nge{#1}}

\def\each@rg#1#2{{\let\thecsname=#1\expandafter\first@rg#2,\end,}}
\def\first@rg#1,{\thecsname{#1}\apply@rg}
\def\apply@rg#1,{\ifx\end#1\let\next=\relax%
\else,\thecsname{#1}\let\next=\apply@rg\fi\next}

\def\callr@nge#1{\calldor@nge#1-\end-}
\def\callr@ngeat#1\end-{#1}
\def\calldor@nge#1-#2-{\ifx\end#2\@qneatspace#1 %
  \else\calll@@p{#1}{#2}\callr@ngeat\fi}
\def\calll@@p#1#2{\ifnum#1>#2{\@rrwrite{Equation range #1-#2\space is bad.}
\errhelp{If you call a series of equations by the notation M-N, then M and
N must be integers, and N must be greater than or equal to M.}}\else%
 {\count0=#1\count1=#2\advance\count1 by1\relax\expandafter\@qncall\the\count0,
  \loop\advance\count0 by1\relax%
    \ifnum\count0<\count1,\expandafter\@qncall\the\count0,%
  \repeat}\fi}

\def\@qneatspace#1#2 {\@qncall#1#2,}
\def\@qncall#1,{\ifunc@lled{#1}{\def\next{#1}\ifx\next\empty\else
  \w@rnwrite{Equation number \noexpand\(>>#1<<) has not been defined yet.}
  >>#1<<\fi}\else\csname @qnnum#1\endcsname\fi}

\let\eqnono=\eqno
\def\eqno(#1){\tag#1}
\def\tag#1$${\eqnono(\displayt@g#1 )$$}

\def\aligntag#1\endaligntag
  $${\gdef\tag##1\\{&(##1 )\cr}\eqalignno{#1\\}$$
  \gdef\tag##1$${\eqnono(\displayt@g##1 )$$}}

\let\eqalignnono=\eqalignno

\def\eqalignno#1{\displ@y \tabskip\centering
  \halign to\displaywidth{\hfil$\displaystyle{##}$\tabskip\z@skip
    &$\displaystyle{{}##}$\hfil\tabskip\centering
    &\llap{$\displayt@gpar##$}\tabskip\z@skip\crcr
    #1\crcr}}

\def\displayt@gpar(#1){(\displayt@g#1 )}

\def\displayt@g#1 {\rm\ifunc@lled{#1}\global\advance\tagnumber by1
        {\def\next{#1}\ifx\next\empty\else\expandafter
        \xdef\csname @qnnum#1\endcsname{\t@ghead\number\tagnumber}\fi}%
  \writenew@qn{#1}\t@ghead\number\tagnumber\else
        {\edef\next{\t@ghead\number\tagnumber}%
        \expandafter\ifx\csname @qnnum#1\endcsname\next\else
        \w@rnwrite{Equation \noexpand\tag{#1} is a duplicate number.}\fi}%
  \csname @qnnum#1\endcsname\fi}

\def\ifunc@lled#1{\expandafter\ifx\csname @qnnum#1\endcsname\relax}

\let\@qnend=\end\gdef\end{\if@qnfile
\immediate\write16{Equation numbers written on []\jobname.EQN.}\fi\@qnend}

\catcode`@=12


\topglue 0.23in
\centerline{\bf LOGARITHMICALLY SLOW COARSENING}
\centerline{\bf IN NONRANDOMLY FRUSTRATED MODELS}
\bigskip
\centerline{Joel D. Shore, James P. Sethna, Mark Holzer, and Veit Elser}
\centerline{Physics Department, Cornell University, Ithaca, New York 14853}

\bigskip
\centerline{ABSTRACT}
\bigskip

We study the growth (``coarsening'') of domains following a quench in an
Ising model with weak
next--nearest--neighbor antiferromagnetic (AFM) bonds and single--spin--flip
dynamics. The AFM bonds introduce free energy
barriers to coarsening and thus greatly slow the dynamics.
In three dimensions, simple physical arguments suggest that the
barriers are proportional to the characteristic length scale
$L(t)$ for quenches below the corner rounding transition
temperature $T_{CR}$.  This should lead to $L(t)\sim\log(t)$ at long times $t$.
Monte Carlo simulations provide strong support for this claim.

We also predict logarithmic growth in a purely two--dimensional tiling
model, which can be thought of as describing a single
interface in our three--dimensional model viewed from the [111] direction.
Here, the slow coarsening dynamics should persist all the way up to the
order--disorder transition (at $T_{CR}$).
However, if (instead of quenching) we cool the model slowly at a rate
$\Gamma$ through $T_{CR}$, the final length scale should have
power--law, not logarithmic, dependence on $1/\Gamma$.
Simulations support both of these claims, which should in
principle be experimentally testable for a [111] interface of sodium chloride.

\bigskip
\centerline{INTRODUCTION}
\bigskip

When a system is quenched from high temperatures to a temperature below the
order--disorder transition, domains form and coarsen.  Of particular
interest is how the characteristic length scale $L(t)$ grows with
time $t$ at long times.

Historically, there have been some theoretical predictions that certain
systems without randomness in their Hamiltonians would show
logarithmically slow coarsening at long times.\refto{PredictionsOfLogs}
For a while, such claims could
not be disproved since the numerical evidence was
ambiguous due to long time transients and finite--size effects.
However, large Monte Carlo simulations,
bolstered by more careful theoretical arguments, eventually showed that
the long time growth in these models obeys the naively--expected
power laws:  $L(t) \sim t^n$
with $n = 1/3$ or $1/2$ (depending on whether the dynamics does or does not
conserve the order parameter, respectively).\refto{NoLogs}
Indeed, the only models known to exhibit
logarithmic domain growth are those which contain randomness explicitly in
their Hamiltonians, such as the random--field Ising model and
spin glasses.

In light of these results, there
seems to be a growing belief that, for nonrandom systems quenched to nonzero
temperature, the $n = 1/3$ and $n = 1/2$ power law behavior is universal
(\ie independent of the details of the Hamiltonian), and even independent
of the dimensionality.
Motivated by the slow dynamics present in glasses,\refto{OurGlassPaper}
we have been looking for counter examples, \ie
models without randomness
which display logarithmically slow ordering dynamics.\refto{MyThesis}

\bigskip
\centerline{ARGUMENT FOR LOGARITHMICALLY SLOW GROWTH}
\bigskip

Consider
the nearest--neighbor Ising ferromagnet on a square or cubic lattice
in $d = 2$ or $3$
dimensions, with frustration added by introducing weak next--nearest--neighbor
(NNN) antiferromagnetic (AFM) bonds.  The Hamiltonian is
$$
        H = -J_1 {\sum_{\rm NN} s_i s_j} + J_2 {\sum_{\rm NNN} s_i s_j}
        \ ,
        \eqno(hamiltonian)
$$
where $s_i = \pm 1$. The first sum is over
all nearest--neighbor (NN) bonds while the second is over all
NNN bonds.  We have chosen our sign convention so that
both $J_1$ and $J_2$ are positive when the NN bonds are
ferromagnetic and the NNN bonds are antiferromagnetic.  We will require
that $J_1 / J_2 > 2 (d - 1)$ so that the ground state for this model is
ferromagnetic.  We will study this Hamiltonian under
single--spin--flip (\ie nonconserved) dynamics.

The NNN AFM bonds introduce free energy
barriers to coarsening and thus greatly slow the dynamics (freezing the system
completely at $T = 0$).
In two dimensions, these barriers are independent of the characteristic length
scale $L(t)$, and thus $L(t) \sim t^{1/2}$ at long times.\refto{MyThesis}
Let us now study what happens in three dimensions by considering the time to
shrink a cubic domain of, say, up spins in a larger sea of down spins (see
Fig.\ 1).  The energy barrier to flip a corner spin (dark gray)
is $12 J_2$.  Once a corner flips, the neighboring spins along an
edge (light gray) can flip in turn, but there is an energy barrier of $4 J_2$
for each to flip.  The barrier to flip the spins along an entire edge is then
$E = 4 J_2 (L+1)$, where $L$ is the linear size of the domain.  The time $t$ to
do this is given by activation over this barrier and is thus
exponential in $L$:
$$
	t = \> \tau_0 \> e^{4 (L + 1) J_2 / T}
	\ .
	\eqno(Activation)
$$
Naively inverting this equation to solve for the size of
the smallest structure which we expect to remain in a coarsening system at
time $t$, we find
$$
        L(t) \sim {T \over{4 J_2}} \log(t/\tau_0)
        \ .
        \eqno(OurLogarithmicGrowth)
$$
This gives the expected result\refto{Lai} that energy barriers which
diverge with the characteristic length scale $L(t)$ should lead to
logarithmically slow coarsening.\refto{BetterEquation}

Of course, the above discussion is only valid in the limit $T \to 0$.
What happens at nonzero temperatures where we must consider not energy
barriers but, rather, {\it free} energy barriers?  Fig.\ 2 shows Monte
Carlo simulation results for the average time $t$
to flip all the spins along the edge of a
cubic domain.  We see that the slope on this Arrhenius plot increases with
domain size $L$, thus confirming our prediction of an activation barrier
which grows with $L$.

Furthermore, if we write
$$
        t = \tau_0(T) e^{F_B(L,T)/T}
       \ ,
       \eqno (Shrink3dForm)
$$
we can perform
a low temperature expansion for both $\tau_0(T)$ and
$F_B(L,T)$.\refto{MyThesis}  The
curves in Fig.\ 2 show the resulting prediction, which has no free parameters
and is in excellent agreement with the simulation results at low temperatures.

We expect our argument for logarithmically slow coarsening to break down
when the free energy barrier per unit length (to flip the spins along
a cube edge) goes to zero.  This occurs at
the corner rounding temperature $T_{CR}$, which
has previously been studied in the context of equilibrium crystal
shapes.\refto{RottmanAndWortis}  In the limit $J_1/J_2 \to \infty$, $T_{CR}$
can be calculated exactly\refto{ShiAndWortis,MyThesis} and yields
$T_{CR} \approx 7.11 J_2$.

\bigskip
\centerline{SIMULATIONS OF THE COARSENING PROCESS}
\bigskip

There is still a large gap in our argument:
Although we have identified a special configuration
in which there are energy barriers that scale with the length scale $L$,
we have not shown that during
the process of coarsening the system will necessarily find itself in
configurations in which it will have to cross these barriers in order to
coarsen further.  It is conceivable that the system could find a path
through configuration space  which goes around
these barriers.  To construct a proof that the barriers must be crossed is
very difficult since it requires a detailed understanding
of the spin configurations which form in a quench.  Instead, we turn to
numerical simulations of the coarsening process in order
to test our conjecture.

Fig.\ 3 shows the growth of $L(t)$ following a quench from infinite temperature
(a random spin configuration) to a final temperature $T$.  Since this is a
log--log plot, power law behavior would give a straight line.  We see that as
the ratio of $T/J_2$ is decreased, the coarsening slows dramatically.
Furthermore, at temperatures below $T_{CR}$ (for $T/J_2 = 2$, 3, and 4),
the Monte Carlo data show some downward curvature on this log--log plot at
late times.
[By contrast, for $J_2 = 0$ and $T/J_2 = 8$, there is no downward curvature
until finite--size effects lead to a sharp leveling off of $L(t)$ once it
is approximately $1/3$ the system size.]
This suggests that the growth is becoming slower than a power law.  To test
whether the growth is compatible with a logarithm at these late times,
we show (solid curves) two--parameter fits to the form
$$
        L(t) = a \log(t/t_0)
       \ ,
       \eqno (LogForm)
$$
over the last two to three decades in time.  We see that although there is
still some systematic deviation, the fits are quite good and are
certainly superior to straight line (power law) fits over the same range.

\bigskip
\centerline{THE TILING MODEL}
\bigskip

We now briefly discuss a closely related model which is also expected to show
logarithmic coarsening.  This is a two--dimensional model for a
single interface in the three--dimensional model as viewed from the [111]
direction.
If we require that the configurations of this interface have no bubbles or
overhangs (when viewed from this direction), then we obtain the so--called
``[111]--restricted solid--on--solid (RSOS) model'' for our three--dimensional
model.\refto{ShiAndWortis} The RSOS restriction corresponds to taking
the limit $J_1/J_2 \to \infty$ in the 3-d model.
Any configuration in the RSOS model (of which an example is shown in Fig.\ 4)
can be represented as a tiling of the plane by $60^\circ$ rhombi of three
different orientations. (The model also has a third representation as an Ising
spin system on a triangular lattice.\refto{ShiAndWortis})

The order--disorder transition in this model occurs at $T_{CR}$.
Above $T_{CR}$, the
interface is rough (\ie the tiles intermingle); below $T_{CR}$, the
interface forms a sharp corner (\ie the tiles phase separate).  When the
system is quenched from infinite temperature to $T \le T_{CR}$, we expect
that the interface will coarsen under the dynamics, which consists of adding or
removing cubes subject to the RSOS restriction.  Since this dynamics
conserves the
order parameter in this model, the naively--expected behavior would be
$L(t) \sim t^{1/3}$.  However, the
mechanism by which the interface coarsens involves activation
over precisely the same sort of barriers which grow with
$L(t)$ as in the three--dimensional model.  Thus,
the same arguments we made
for logarithmically slow coarsening in the three--dimensional model should
apply here as well.
Simulations of the coarsening process once again lend support to this claim.

Unlike in the three--dimensional model, here the ordering temperature and
the temperature at which the dynamics becomes slow coincide.  Thus, we might
hope that this system would be glassy, \ie  that it would
have great difficultly ordering even when cooled
slowly at a rate $\Gamma$.
Specifically, we'd want the final ($T =0$) value of $L$ to depend
only logarithmically on the time $1/\Gamma$ spent cooling.
We have simulated slow cooling in this
model and find that this does not appear to be the case.  Furthermore, more
careful
arguments suggest that we should expect $L(T = 0) \sim \Gamma^r$  with
$r = -1/4$ in the limit $\Gamma \to 0$, which is
in reasonably good agreement with the simulation results.\refto{MyThesis}
The reason why the dependence is a power
law and not a logarithm is because the free energy barrier goes continuously
to zero at $T_{CR}$, and thus there is a region of temperature just below
$T_{CR}$ where the barriers are small and the system can still coarsen quite
rapidly.  Note, however, that the growing barriers do modify the exponent $r$
from what we would expect in their absence ($r = -1/3$).


\bigskip
\centerline{CONCLUSIONS}
\bigskip

We have discussed two closely related models in which
we conjecture that the growth of the domains should be
only logarithmic in time following a quench.  Simulations lend strong support
to this conjecture.  However, if cooled slowly at a rate $\Gamma$ (rather than
quenched), these models are not expected to order sluggishly: the final
domain size has a power law, rather than a logarithmic, dependence on
$1/\Gamma$.

\bigskip
\centerline{ACKNOWLEDGEMENTS}
\bigskip

We thank David Huse, Peter Nightingale, Jennifer Hodgdon,
and David DiVincenzo for helpful discussions.
This work was supported in part by NSF Grant No. DMR 88-15685 and computing
facilities were provided in part by the Cornell--IBM Joint Study on Computing
for Scientific Research.

\bigskip
\centerline{REFERENCES}
\bigskip

   \references

\refis{PredictionsOfLogs} S.~A. Safran, Phys. Rev. Lett. {\bf 46}, 1581 (1981);
G.~F. Mazenko, O.~T. Valls, and F.~C. Zhang, Phys. Rev. B {\bf 31}, 4453
(1985).

\refis{OurGlassPaper}
For more details, see J.~P. Sethna, J.~D. Shore, and M. Huang,
\prb 44, 4943, 1991; and references therein.


\refis{RottmanAndWortis}
C. Rottman and M. Wortis, \prb 29, 328, 1984.

\refis{Lai} Z.~W. Lai, G.~F. Mazenko, and O.~T. Valls, \prb 37, 9481, 1988.

\refis{MyThesis} More details on the work presented here can be found in
J.~D. Shore, Mark Holzer, and J.~P. Sethna, submitted to {\sl Phys. Rev. B}
(Electronic bulletin board \#9204015).
An earlier report on some of the work is given in J.~D. Shore and
J.~P. Sethna, \prb 43, 3782, 1991.

\refis{ShiAndWortis}
A.-C. Shi and M. Wortis, \prb 37, 7793, 1988; and references therein.

\refis{NoLogs} D.~A. Huse, \prb 34, 7845, 1986;
J. Vi\~nals and M. Grant, \prb 36, 7036, 1987;
G.~S. Grest, M.~P. Anderson, and D.~J. Srolovitz, \prb 38, 4752, 1988;
J.~G. Amar, F.~E. Sullivan, and R.~D. Mountain, \prb 37, 196, 1988;
C. Roland and M. Grant, \prb 39, 11971, 1989.

\refis{BetterEquation}  More precisely, the differential
equation satisfied at late times in the coarsening process should be of the
form
$$
        {dL\over dt} \sim {e^{-4 J_2 L /T} \over{L}}
       \ .
        \eqno(TrueEquation)
$$
Such an equation expresses the fact that we still have curvature--driven
growth, but with a prefactor reflecting activation over free energy
barriers proportional to $L$.  In the limit $t \to \infty$,
Eq.\ \(TrueEquation) yields Eq.\ \(OurLogarithmicGrowth).

\endreferences

\vfill\eject


\catcode`\@=11\relax
\newwrite\@unused
\def\typeout#1{{\let\protect\string\immediate\write\@unused{#1}}}
\typeout{psfig/tex 1.2-dvi2ps-li}

\def\figurepath{./}
\def\psfigurepath#1{\edef\figurepath{#1}}

\def\@nnil{\@nil}
\def\@empty{}
\def\@psdonoop#1\@@#2#3{}
\def\@psdo#1:=#2\do#3{\edef\@psdotmp{#2}\ifx\@psdotmp\@empty \else
    \expandafter\@psdoloop#2,\@nil,\@nil\@@#1{#3}\fi}
\def\@psdoloop#1,#2,#3\@@#4#5{\def#4{#1}\ifx #4\@nnil \else
       #5\def#4{#2}\ifx #4\@nnil \else#5\@ipsdoloop #3\@@#4{#5}\fi\fi}
\def\@ipsdoloop#1,#2\@@#3#4{\def#3{#1}\ifx #3\@nnil
       \let\@nextwhile=\@psdonoop \else
      #4\relax\let\@nextwhile=\@ipsdoloop\fi\@nextwhile#2\@@#3{#4}}
\def\@tpsdo#1:=#2\do#3{\xdef\@psdotmp{#2}\ifx\@psdotmp\@empty \else
    \@tpsdoloop#2\@nil\@nil\@@#1{#3}\fi}
\def\@tpsdoloop#1#2\@@#3#4{\def#3{#1}\ifx #3\@nnil
       \let\@nextwhile=\@psdonoop \else
      #4\relax\let\@nextwhile=\@tpsdoloop\fi\@nextwhile#2\@@#3{#4}}

\def\psdraft{
	\def\@psdraft{0}
}
\def\psfull{
	\def\@psdraft{100}
}
\psfull
\newif\if@prologfile
\newif\if@postlogfile
\newif\if@noisy
\def\pssilent{
	\@noisyfalse
}
\def\psnoisy{
	\@noisytrue
}
\psnoisy
\newif\if@bbllx
\newif\if@bblly
\newif\if@bburx
\newif\if@bbury
\newif\if@height
\newif\if@width
\newif\if@rheight
\newif\if@rwidth
\newif\if@clip
\newif\if@verbose
\def\@p@@sclip#1{\@cliptrue}

\def\@p@@sfile#1{\def\@p@sfile{null}%
	        \openin1=#1
		\ifeof1\closein1%
		       \openin1=\figurepath#1
			\ifeof1\typeout{Error, File #1 not found}
			\else\closein1
			    \edef\@p@sfile{\figurepath#1}%
                        \fi%
		 \else\closein1%
		       \def\@p@sfile{#1}%
		 \fi}
\def\@p@@sfigure#1{\def\@p@sfile{null}%
	        \openin1=#1
		\ifeof1\closein1%
		       \openin1=\figurepath#1
			\ifeof1\typeout{Error, File #1 not found}
			\else\closein1
			    \def\@p@sfile{\figurepath#1}%
                        \fi%
		 \else\closein1%
		       \def\@p@sfile{#1}%
		 \fi}

\def\@p@@sbbllx#1{
		\@bbllxtrue
		\dimen100=#1
		\edef\@p@sbbllx{\number\dimen100}
}
\def\@p@@sbblly#1{
		\@bbllytrue
		\dimen100=#1
		\edef\@p@sbblly{\number\dimen100}
}
\def\@p@@sbburx#1{
		\@bburxtrue
		\dimen100=#1
		\edef\@p@sbburx{\number\dimen100}
}
\def\@p@@sbbury#1{
		\@bburytrue
		\dimen100=#1
		\edef\@p@sbbury{\number\dimen100}
}
\def\@p@@sheight#1{
		\@heighttrue
		\dimen100=#1
   		\edef\@p@sheight{\number\dimen100}
}
\def\@p@@swidth#1{
		\@widthtrue
		\dimen100=#1
		\edef\@p@swidth{\number\dimen100}
}
\def\@p@@srheight#1{
		\@rheighttrue
		\dimen100=#1
		\edef\@p@srheight{\number\dimen100}
}
\def\@p@@srwidth#1{
		\@rwidthtrue
		\dimen100=#1
		\edef\@p@srwidth{\number\dimen100}
}
\def\@p@@ssilent#1{
		\@verbosefalse
}
\def\@p@@sprolog#1{\@prologfiletrue\def\@prologfileval{#1}}
\def\@p@@spostlog#1{\@postlogfiletrue\def\@postlogfileval{#1}}
\def\@cs@name#1{\csname #1\endcsname}
\def\@setparms#1=#2,{\@cs@name{@p@@s#1}{#2}}

\def\ps@init@parms{
		\@bbllxfalse \@bbllyfalse
		\@bburxfalse \@bburyfalse
		\@heightfalse \@widthfalse
		\@rheightfalse \@rwidthfalse
		\def\@p@sbbllx{}\def\@p@sbblly{}
		\def\@p@sbburx{}\def\@p@sbbury{}
		\def\@p@sheight{}\def\@p@swidth{}
		\def\@p@srheight{}\def\@p@srwidth{}
		\def\@p@sfile{}
		\def\@p@scost{10}
		\def\@sc{}
		\@prologfilefalse
		\@postlogfilefalse
		\@clipfalse
		\if@noisy
			\@verbosetrue
		\else
			\@verbosefalse
		\fi

}

\def\parse@ps@parms#1{
	 	\@psdo\@psfiga:=#1\do
		   {\expandafter\@setparms\@psfiga,}}

\newif\ifno@bb
\newif\ifnot@eof
\newread\ps@stream
\def\bb@missing{
	\if@verbose{
		\typeout{psfig: searching \@p@sfile \space  for bounding box}
	}\fi
	\openin\ps@stream=\@p@sfile
	\no@bbtrue
	\not@eoftrue
	\catcode`\%=12
	\loop
		\read\ps@stream to \line@in
		\global\toks200=\expandafter{\line@in}
		\ifeof\ps@stream \not@eoffalse \fi
		\@bbtest{\toks200}
		\if@bbmatch\not@eoffalse\expandafter\bb@cull\the\toks200\fi
	\ifnot@eof \repeat
	\catcode`\%=14
}
\catcode`\%=12
\newif\if@bbmatch
\def\@bbtest#1{\expandafter\@a@\the#1
\long\def\@a@#1
	\else\@bbmatchtrue\fi}
\long\def\bb@cull#1 #2 #3 #4 #5 {
	\dimen100=#2 bp\edef\@p@sbbllx{\number\dimen100}
	\dimen100=#3 bp\edef\@p@sbblly{\number\dimen100}
	\dimen100=#4 bp\edef\@p@sbburx{\number\dimen100}
	\dimen100=#5 bp\edef\@p@sbbury{\number\dimen100}
	\no@bbfalse
}
\catcode`\%=14
\def\compute@bb{
		\no@bbfalse
		\if@bbllx \else \no@bbtrue \fi
		\if@bblly \else \no@bbtrue \fi
		\if@bburx \else \no@bbtrue \fi
		\if@bbury \else \no@bbtrue \fi
		\ifno@bb \bb@missing \fi
		\ifno@bb \typeout{FATAL ERROR: no bb supplied or found}
			\no-bb-error
		\fi
		\count203=\@p@sbburx
		\count204=\@p@sbbury
		\advance\count203 by -\@p@sbbllx
		\advance\count204 by -\@p@sbblly
		\edef\@bbw{\number\count203}
		\edef\@bbh{\number\count204}
}

\def\in@hundreds#1#2#3{\count240=#2 \count241=#3
		     \count100=\count240	
		     \divide\count100 by \count241
		     \count101=\count100
		     \multiply\count101 by \count241
		     \advance\count240 by -\count101
		     \multiply\count240 by 10
		     \count101=\count240	
		     \divide\count101 by \count241
		     \count102=\count101
		     \multiply\count102 by \count241
		     \advance\count240 by -\count102
		     \multiply\count240 by 10
		     \count102=\count240	
		     \divide\count102 by \count241
		     \count200=#1\count205=0
		     \count201=\count200
			\multiply\count201 by \count100
		 	\advance\count205 by \count201
		     \count201=\count200
			\divide\count201 by 10
			\multiply\count201 by \count101
			\advance\count205 by \count201
		     \count201=\count200
			\divide\count201 by 100
			\multiply\count201 by \count102
			\advance\count205 by \count201
		     \edef\@result{\number\count205}
}
\def\compute@wfromh{
		\in@hundreds{\@p@sheight}{\@bbw}{\@bbh}
		\edef\@p@swidth{\@result}
}
\def\compute@hfromw{
		\in@hundreds{\@p@swidth}{\@bbh}{\@bbw}
		\edef\@p@sheight{\@result}
}
\def\compute@handw{
		\if@height
			\if@width
			\else
				\compute@wfromh
			\fi
		\else
			\if@width
				\compute@hfromw
			\else
				\edef\@p@sheight{\@bbh}
				\edef\@p@swidth{\@bbw}
			\fi
		\fi
}
\def\compute@resv{
		\if@rheight \else \edef\@p@srheight{\@p@sheight} \fi
		\if@rwidth \else \edef\@p@srwidth{\@p@swidth} \fi
}

\def\compute@sizes{
	\compute@bb
	\compute@handw
	\compute@resv
}

\def\psfig#1{\vbox {
	\ps@init@parms
	\parse@ps@parms{#1}
	\compute@sizes
	\ifnum\@p@scost<\@psdraft{
		\if@verbose{
			\typeout{psfig: including \@p@sfile \space }
		}\fi
		\special{ pstext="\@p@swidth \space
			\@p@sheight \space
			\@p@sbbllx \space \@p@sbblly \space
			\@p@sbburx  \space
			\@p@sbbury \space startTexFig" \space}
		\if@clip{
			\if@verbose{
				\typeout{(clip)}
			}\fi
			\special{ pstext="doclip \space"}
		}\fi
		\if@prologfile
		    \includegraphics{\@prologfileval} \fi
		\includegraphics{\@p@sfile}
		\if@postlogfile
		    \includegraphics{\@postlogfileval} \fi
		\special{pstext=endTexFig \space }
		\vbox to \@p@srheight true sp{
			\hbox to \@p@srwidth true sp{
				\hss
			}
			\vss
		}
	}\else{
		\vbox to \@p@srheight true sp{
		\vss
			\hbox to \@p@srwidth true sp{
				\hss
				\if@verbose{
					\@p@sfile
				}\fi
				\hss
			}
		\vss
		}
	}\fi
}}
\def\psglobal{\typeout{psfig: PSGLOBAL is OBSOLETE; use psprint -m instead}}
\catcode`\@=12\relax

\def\boxit#1{\vbox{\hrule\hbox{\vrule\vbox{#1}\vrule}\hrule}}
\let\PsfiG\psfig
\let\PsdrafT\psdraft
\let\PsfulL\psfull
\def\psfig#1{\BoxiT{\PsfiG{#1}}}
\def\BoxiT{\relax}
\def\psdraft{\def\BoxiT{\boxit}\PsdrafT}
\def\psfull{\def\BoxiT{\relax}\PsfulL}


\topinsert
\centerline{\psfig{figure=cube.ps,width=6.5truecm}}
\smallskip
\endinsert

\centerline{\vbox{
\hsize= 6.0truecm
{\bf Fig 1.} A cubic domain of size $L$ of up spins in a larger sea of down
spins.
}
}

\vfill\eject

\topinsert
\centerline{\psfig{figure=shrinking_cubes.ps,width=15.0truecm}}
\smallskip
\endinsert

\centerline{\vbox{
\hsize= 12.0truecm
{\bf Fig 2.} Arrhenius plot of the time to flip all the spins
along the edge of a cubic domain of size $L$ (shown in Fig.\ 1).
Curves are the theoretical forms derived from a low temperature
expansion, as dis\-cussed in the text.
}
}

\vfill\eject

\topinsert
\centerline{\psfig{figure=coarsening.ps,width=15.0truecm}}
\smallskip
\endinsert
\centerline{\vbox{
\hsize= 12.0truecm
{\bf Fig 3.} Growth of $L(t)$ following a quench from infinite
temperature to a final temperature $T$.  Numbers in parentheses
give system sizes.  The solid curves show two--parameter fits
(over the interval for which the curve is shown) to the form
$L(t) = a \, \log(t/t_0)$.
Arrows for $T = 0.75 J_2$ indicate the times at which
$t = e^{4J_2/T}$ and $e^{8J_2/T}$.
}
}

\vfill\eject

\topinsert
\centerline{\psfig{figure=still_12.ps,width=14.0truecm}}
\smallskip
\endinsert

\centerline{\vbox{
\hsize= 10.0truecm
{\bf Fig 4.} A sample configuration for the tiling model.
}
}

\vfill\eject

\end